\documentclass[a4paper]{article}

\usepackage[english]{babel}
\usepackage[utf8x]{inputenc}
\usepackage[T1]{fontenc}
\usepackage{authblk}

\usepackage[a4paper,top=3cm,bottom=2cm,left=3cm,right=3cm,marginparwidth=1.75cm]{geometry}

\usepackage{amsmath}
\usepackage{graphicx}
\usepackage[superscript,biblabel]{cite}
\usepackage[colorinlistoftodos]{todonotes}
\usepackage[colorlinks=true, allcolors=blue]{hyperref}

\title{Contractility in an Extensile System}
\author[1]{Kasimira T. Stanhope}
\author[1]{Vikrant Yadav}
\author[1]{Christian D. Santangelo}
\author[1]{Jennifer L. Ross*}
\affil[1]{Department of Physics, University Massachusetts Amherst, 01003}
\date{}

\begin{document}
\maketitle

\begin{abstract}
Essentially all biology is active and dynamic. Biological entities autonomously sense, compute, and respond using energy-coupled ratchets that can produce force and do work. The cytoskeleton, along with its associated proteins and motors, is a canonical example of biological active matter, which is responsible for cargo transport, cell motility, division, and morphology. Prior work on cytoskeletal active matter systems showed either extensile or contractile dynamics. Here, we demonstrate a cytoskeletal system that can control the direction of the network dynamics to be either extensile, contractile, or static depending on the concentration of filaments or transient crosslinkers through systematic variation of the crosslinker or microtubule concentrations. Based off these new observations and our previously published results, we created a simple one-dimensional model of the interaction of filaments within a bundle. Despite its simplicity, our model recapitulates the observed activities of our experimental system, implying that the dynamics of our finite networks of bundles are driven by the local filament-filament interactions within the bundle. Finally, we show that contractile phases can result in autonomously motile networks that resemble cells.  Our experiments and model allow us to gain a deeper understanding of cytoskeletal dynamics and provide a stepping stone for designing active, autonomous systems that could potentially dynamically switch states.
\end{abstract}

\section*{Introduction}
Active matter appears in biology in various forms. Examples range in scale from intracellular transport to cell locomotion and from bacterial turbulence to swarms of birds. These non-equilibrium, driven  systems  show dynamics that are unlikely or impossible in passive, thermodynamically-equilibrated systems. This makes it profitable to adopt an active matter approach for understanding the complex dynamics of biological systems. Cytoskeletal networks are some of the most important active, mechano-chemical systems in nature. Composed of microtubules, actin, and intermediate filaments, the cytoskeleton utilizes an array of crosslinkers and molecular motors to self-organize  the cellular interior. It is responsible for intracellular cargo transport, cell division, locomotion, and cell shape, yet we do not fully understand how the network is actively reorganized to facilitate these activities. One reason for this lack of understanding is the large number of constituent components in the cytoskeleton and the complexity of their interactions. In this work, we study the dynamics of a simplified, reconstituted cytoskeleton: microtubules interacting via weak, transient crosslinkers and driven by molecular motors. We show that this system is capable of reproducing extensile, contractile, and static dynamic phases reminiscent of those observed in cells.

Microtubules inside the cell need to rearrange themselves during different developmental and cell division stages. During cell division, microtubules change from an aster-like configuration during interphase to the bipolar mitotic spindle that aligns the chromosomes and then separates and pushes apart chromosomes during cell division and cytokinesis. These highly anisotropic structures are formed by bundling microtubules with microtubule associated proteins (MAPs). MAP65 is one such protein used  to control microtubule networks in mitosis and the plant cell cortex to form antiparallel microtubule arrays. MAP65 is a plant protein, but has analogs in yeast (Ase1) and humans (PRC1) \cite{Braun2011,Loiodice2005,Bieling2010,Subramanian2010,Lansky2015}. It acts by dimerizing and bridging between two microtubules to form a microtubule bundle with antiparallel arrangement and space of 35 nm between the filaments \cite{Jiang1998,Loiodice2005,Subramanian2010,Chan1999}. We have shown that the bundles observed at low MAP65 and low microtubule concentrations become networks of bundles when either are increased \cite{Pringle2013}. Further, individual MAP65 are transient binders to single microtubules \cite{Pringle2013}, and the affinity is relatively low in dilute solution, 1.2 $\mu$M \cite{Tulin2012}. Despite this low affinity and rapid turn-over, we showed that MAP65 binds rapidly to antiparallel microtubules and can stall microtubules gliding under the power of kinesin-1 motor proteins \cite{Pringle2013}.

Kinesin-1 is a vital enzymatic microtubule-associated motor protein responsible for long range cargo transport inside cells \cite{Hendricks2010,Vale2000,Hirokawa2009}. When fixed to a substrate, kinesin-1 can be used to drive microtubule gliding, fueled by ATP hydrolysis. We and other groups have routinely used this experimental platform to investigate self-organization of propelled filamentous particles \cite{Liu2011,Pringle2013,Sumino2012,Surrey2010}. Similar studies have been performed using myosin-II propelling actin filaments \cite{Schaller2010}. The facility of such filament gliding assays have resulted in them being engineered in lab-on-a-chip devices to produce nanoscale transport processes, termed "molecular shuttles"  \cite{Vogel2001,Vogel2006,Vogel2007}. 

Other active cytoskeletal systems use crosslinked motors (kinesins or myosins) to drive the motion of cytoskeletal filaments and examine network steady-states \cite{Leibler1997,Sanchez2012, Alvarado2013,Murrell2012,Koenderink2009,Stachowiak2012}. Most of these steady-states for actin-myosin result in contracting networks. Microtubule-kinesin networks have proven more interesting creating steady-state asters \cite{Leibler1997} or dynamically extensile active liquid crystals or cilia-like beating \cite{Sanchez2012,Sanchez2013}. A microtubule, unlike any other filamentous component of the cytoskeleton, has a enormous persistence length, $1~$mm \cite{Hawkins2010}. Thus, $10-30 ~\mu m$ long microtubules behave, essentially, as rigid rods. Unlike actomyosin networks, this stiffness tends to lead to alignment within bundles and, consequently, extensile motion under motor driving \cite{Sanchez2012}.

Contractile steady-states of microtubules driven by crosslinking motors typically evolve from asters \cite{Leibler1997,Needleman2015,Nedelec2016,Kazuhiro2016}. Some of these experiments were conducted in cell extracts making it unclear which cellular components are responsible for the novel contractile phases \cite{Needleman2015}. Other experiments in vitro by different groups using the same components have reported contradictory results \cite{Kazuhiro2016,Surrey2010}. Together, prior work tells a story that asters are needed for contraction and bundles are extensile. Further, in all previous studies, the system evolves to extension or contraction exclusively without room for tuning or regulating the observed states.

Using the microtubule gliding assay driven by kinesin-1 motors, and a network of transiently-crosslinked bundles associated via MAP65, we have designed an active, crosslinked microtubule network. This minimal system allowed us to vary the relative concentrations of crosslinkers and microtubules independently to examine their effect on the dynamics of the network. Excitingly, we observed a wide variety of dynamical activities, including extension and contraction. The dynamics was dependent on the concentration of crosslinkers or the denisty of filaments, which we varied as control parameters. At low concentrations of crosslinkers or microtubule densities, the network spread out under the driving of kinesin-1 motors. As we increased either the relative concentration of crosslinker or microtubules, the finite networks of bundles became contractile. At the highest concentrations of cross-linkers or microtubule densities, all network dynamics ceased. Each dynamic activity was characterized by measuring the rate of spreading of the finite microtubule networks. The spreading speed was constant over experimental times irrespective of the network's phase, which could have been extensile or contractile. The spreading rate was nonmonotonic as a function of either crosslinker concentration or microtubule density. Finally, we modeled the behavior of networks of bundles using a simplified one-dimensional molecular dynamic simulation. Our model was able to recover the crossover from an extensile to contractile phase that we observed in experiments. Together, our observations and simple one-dimensional model gives new insights on the mechanism driving the dynamics of the two-dimensional networks of bundles, showing that the dynamics are driven by the microtubule-microtubule interactions.  

\section*{Experimental}
Tubuin,  kinesin, and all other reagents and materials were fully described in our prior methods chapter \cite{Stanhope2015}. Also, all methodological details can be found in Stanhope and Ross \cite{Stanhope2015} and Pringle et. al. \cite{Pringle2013}. Briefly, fluorescent microtubules were incubated with MAP65 at a specific concentration (variable) in the presence of 35$\mu M$ Taxol and 35$\mu M$ DTT in PEM-100 buffer (100 mM K-PIPES, 1 mM $MgSO_4$, 1 mM EGTA, pH 6.8) for 30 minutes. We made a 10$\mu L$ flow chamber using a slide, double-stick tape, and a cover slip. The chamber was incubated with kinesin-1 at room temperature for 5 minutes. We flowed BSA (5 $mg/ml$ BSA, 60 $\mu M$ Taxol, 20 $mM$ DTT in PEM-100) through the channel. Prior to insertion of microtubule networks in the chamber, we added a motility mix containing  1 mM ATP, 1.9 $\mu M$ phosphocreatine, 68 $\mu g/ml$ creatin phosphokinase, 4.3 $mg/ml$ glucose, and 0.4 $\mu L$ of deoxygenation system (6 units/ml of glucose oxidase and 0.177 mg/ml of catalase) for every 10 $\mu L$ volume in PEM-100 to the networks.  The network dynamics in the chamber were imaged using epi-fluorecence. We waited approximately 5 minutes to allow the networks to sediment and interact with the kinesin-1 on the cover slip. Images were taken every 20 seconds for a duration of up to 1-3 hours. At least three separate chambers were used for each configuration of microtubule and crosslinker concentrations.

Analysis was performed after networks were in full contact with the surface and in a two-dimensional configuration to ensure that the changes in fluorescence were only due to motion in the imaging plane and not due to filaments coming into the imaging plane from the third dimension. As the microtubule bundle expanded, the decay in signal to noise ratio made tracking the edges difficult. To overcome this problem we use a predictor-corrector type algorithm developed in Matlab (Supplemental Information)\cite{NR}. 
The first step of the algorithm used an edge detection function to find the edges of the network. We tuned the parameters so that only the interior was detected. Next, the interior portion was subtracted from the original image. The corrected image was used as an input for the next round of edge detection. By tuning the sensitivity of edge detection and the number of iterations, we were able to detect edges robustly for all time points.

We repeated the experiments on at least three different networks in different experimental chambers. The normalized variance of the network shape (see below) was averaged together before plotting to smooth variations in microtubule density that might exist between the distinct networks of bundles.

\section*{Results and Discussion}

We created microtubule networks by incubating Taxol stabilized microtubules with MAP65 crosslinkers prior to performing a kinesin-driven filament gliding assay \cite{Pringle2013,Stanhope2015}. We previously measured the equilibrium phases of microtubules with MAP65 crosslinking in the absence of kinesin motor driving. We found either single filaments at low MAP65 concentrations, individual bundles as higher MAP65 concentrations, and low microtubule concentrations, or networks of bundles when the microtubule and MAP65 concentrations were sufficiently large  \cite{Pringle2013}. 

\begin{figure}
	\centering
		\includegraphics[width=1.0\linewidth]{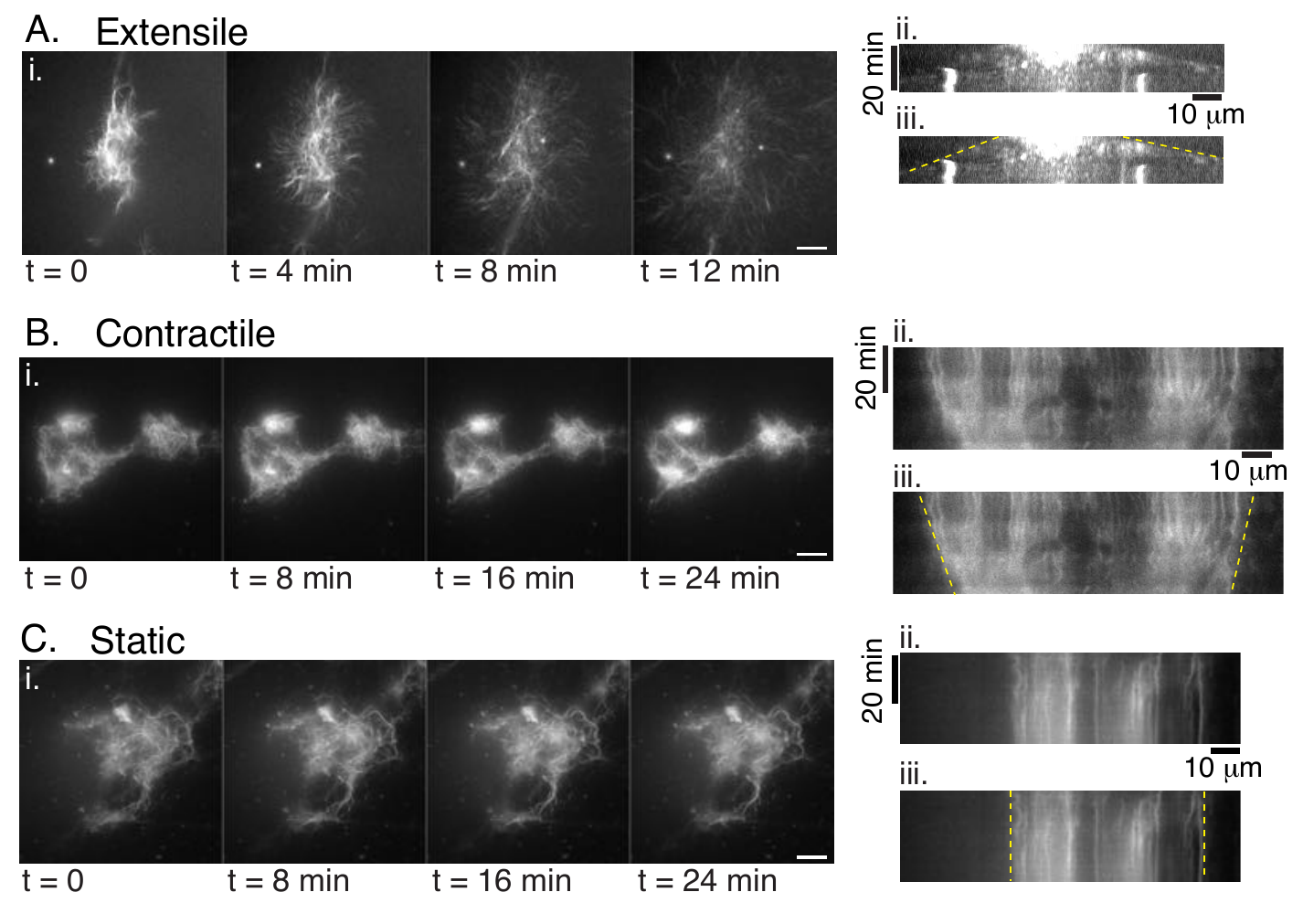}
		\caption{Examples of the different types of activities observed in the active network experiments. (A) Example of an extensile network. (i) Timeseries of extensile network at 4 minute intervals. Microtubule concentration is 2.5 $\mu M$, and MAP65 is at 13\%. Scale bar is 20 $\mu M$. (ii) Kymograph made through the center of the network reveals the motion of the network. Vertical scale bar is 20 min, horizontal scale bar is 10 $\mu M$. (iii) The same kymograph with the edges that are extending marked (dashed, yellow line). (B) Example of a contractile network. Microtubule concentration is 2.5 $\mu M$, and MAP65 is at 35\%. (i) Timeseries of contractile network at 8 minute intervals. Scale bar is 20 $\mu M$. (ii) Kymograph made through the center of the network reveals the motion of the network. Vertical scale bar is 20 min, horizontal scale bar is 10 $\mu M$. (iii) The same kymograph with the edges that are contracting marked (dashed, yellow line). (C) Example of a static network. Microtubule concentration is 10 $\mu M$, and MAP65 is at 13\%. (i) Timeseries of static network at 8 minute intervals. Scale bar is 20 $\mu M$. (ii) Kymograph made through the center of the network reveals the small initial motion of the network and became static. Vertical scale bar is 20 min, horizontal scale bar is 10 $\mu M$. (iii) The same kymograph with the edges that are static marked (dashed, yellow line).}
	\label{fig:Trackfig}
\end{figure}

Here, we used the same networks of bundles in the relatively high MAP65 and microtubule regimes and added them to a kinesin-gliding assay. We found that these networks could dynamically rearrange under the power of the kinesin-1 driving by either extending, contracting, or remaining static, depending on the relative concentration of crosslinkers and microtubules  (Fig. \ref{fig:Trackfig}, Supp. Fig. 1). We quantified the spatial distribution of microtubules in each network over time by tracking the edges of the network from images obtained via fluorescence microscopy (see supplemental information for details on this tracking). At least three or more different timelapse movies of network evolution were analyzed corresponding to each microtubule and crosslinker concentration, totaling to 72 different data sets. 

Once we identified the edges of the network over time, we used those pixel locations to quantify the shape of the network by measuring the second moment $(\sigma)$ of the edge:
\begin{equation}
\sigma = \sqrt{\sum\limits_{i=1}^{N}\frac{(x_i - \mu)^2}{N}}.
\label{Definition}
\end{equation}
Here, $x_i$ is the $i^{th}$ pixel detected on the edge of the network, $N$ is the total number of pixels detected on the edge, and $\mu$ is the center of mass of the network. As each network may have an arbitrary initial shape and size, we normalized $\sigma$ by its value at time $t = 0$, $\sigma_0$. We plot the normalized variance of the shape ($\sigma/\sigma_0$) as a function of time (Fig. \ref{fig:CLMTfig}A,B). We observed that the change in $\sigma/\sigma_0$ was linear with time (Fig. \ref{fig:CLMTfig}A,B). Accordingly, we performed a least-squared fit of the normalized $\sigma$ data \textit{vs.} time to the linear equation:
\begin{equation}
\frac{\sigma_t}{\sigma_0} = \omega t + 1.
\label{FitEquation}
\end{equation}
This equation defined the spreading speed $\omega$ as the slope of the best linear fit. The spreading speed, $\omega$, provides a metric to classify network dynamics as either extensile ($\omega > 0$), contractile ($\omega < 0$), or static ($\omega = 0$).

We will use this metric to determine the type of dynamics of the network as a function of the percentage of crosslinker bound, $\rho_{c}$, (relative to the maximal crosslinker binding) and the concentration of microtubules, $\rho_{MT}$.  It is interesting that we always observed the size of the network of bundles to change linearly with time. If the network of bundles was simply being driven apart by a random, thermal process, we would expect the size to change with the square-root of time, as one might expect for a droplet spreading under diffusion. Further, the linear evolution is in contrast with the subdiffusive dynamics seen in cell spreading experiments \cite{Mahadevan2007}. The linear spreading makes sense because the system is driven by processive motors that move with a constant velocity. Previously, processes of network or cell spreading of this type have been dubbed "active wetting" \cite{Joanny2013}. 

\subsection*{The Dynamics of Driven Networks Depend on Crosslinker and Microtubule Density}

\begin{figure}
	\centering
		\includegraphics[width=1\linewidth]{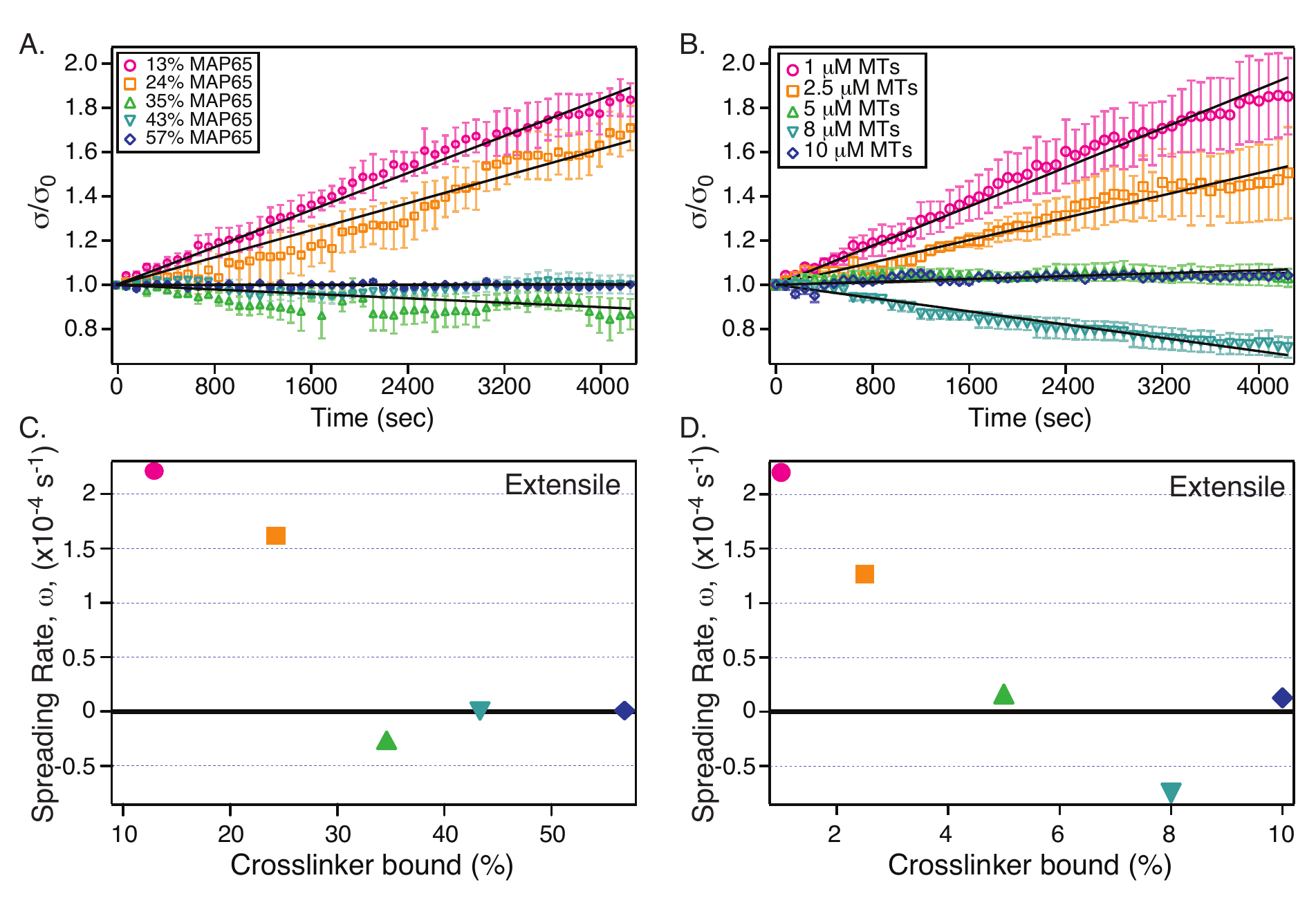}
		\caption{Effect of crosslinker concentration and microtubule density on microtubule network dynamics. (A) Normalized shape variance, $\sigma/\sigma_0$, as a function of time for a constant microtubule concentration of 2.5 $\mu M$ and increasing crosslinker percentage. Each point represents the average of at least three distinct networks. The error bars represent the standard error of the mean. Each line was fit to a line, equation \ref{FitEquation}, and the spreading rate, $\omega$ was determined from the slope. (B) Normalized shape variance, $\sigma/\sigma_0$, as a function of time for a constant binding percentage of 13 \% and increasing microtubule concentration. Each point represents the average of at least three distinct networks. The error bars represent the standard error of the mean. Each line was fit to a line, equation (2), and the spreading rate, $\omega$ was determined from the slope. (C) Spreading rate $(\omega)$ of a microtubule network as a function of percentage of crosslinker bound is non-monotonic due to the network contraction. (D) Spreading rate $(\omega)$ of a microtubule network as a function of microtubule concentration is also non-monotonic due to the network contraction.}
	\label{fig:CLMTfig}
\end{figure}

We performed a series of experiments for a variety of different microtubule concentrations and crosslinker percentages. Specifically, we bundled the microtubules at each concentration (1.0, 2.5, 5.0, 8.0, 10.0 $\mu M$) with a concentration of MAP65 crosslinker that would result in a prescribed percentage of MAP65 bound to the microtubules (13, 24, 35, 43, 57\%). The percentage of crosslinker bound to microtubules was controlled by assuming that the equilibrium binding affinity $K_D =$ 1.2 $\mu M$ as previously measured for dilute microtubule-MAP65 conditions \cite{Tulin2012}. We have previously determined the equilibrium phases for MAP65-crosslinkers with microtubules using this parameter \cite{Pringle2013}. 

As an example, we show the change in the size of the networks for one microtubule concentration (2.5 $\mu M$) (Fig. \ref{fig:CLMTfig}A). At low crosslinker concentrations, the spreading rate, $\omega$, was positive because the network was extensile (Fig. \ref{fig:CLMTfig}C). With increasing crosslinker concentration, the networks slowed down and changed from extensile to contractile motion with a negative spreading rate, $\omega$ (Fig. \ref{fig:CLMTfig}C). At the highest concentration of crosslinker, there were no dynamics, and the spreading rate was zero (Fig. \ref{fig:CLMTfig}C). Together, we observed that the spreading rate in a non-monotonic function of crosslinker density (Fig. \ref{fig:CLMTfig}C).

The dynamics of our networks of bundles system were not only controlled by the concentration of crosslinker, but also by the concentration of microtubules present. We observed the same pattern of extensile, contractile, and static activities when we increased the concentration of microtubules and held the relative crosslinker density fixed at 13\% (Fig. \ref{fig:CLMTfig}B). Further, the same non-monotonic behavior of the spreading rate $\omega$ was likewise observed as the microtubule concentration increased (Fig. \ref{fig:CLMTfig}D). 

In order to examine if the presence of crosslinker is essential to the existence of a contractile phase, we performed a series of control experiments where the microtubules were bundled via depletion interactions using polyethylene glycol (PEG) as a crowding agent. We introduced the PEG-bundles to a kinesin-driven gliding assay. We never observed any contraction in this system. Thus, bundling by itself is not the cause of contractility -- the specific, cohesive crosslinking activity is required. In our previous work, we showed that the MAP65 crosslinking is cooperative and quick acting in the presence of antiparallel microtubules. Further, MAP65 crosslinking was capable of dynamically slowing down microtubule gliding between only two filaments \cite{Pringle2013}. The slower speed was a function of higher crosslinker concentration.

We were intrigued by the contractile activity of the networks, which has never before been observed in gliding microtubule systems. Prior work with actin filament networks showed that contraction was mediated by filament buckling \cite{Murrell2012}. In order to investigate if individual filaments were able to buckle, we used two color experiments to monitor individual tracer microtubules within differentially labeled networks. We never detected microtubule buckling during contractile phases (data not shown). The absence of filament buckling is not surprising given how much stiffer the microtubule filament is, with a persistence length of about 1 mm \cite{Hawkins2010,Hawkins2012,Hawkins2013} compared to the actin, with a persistence length of 10-20 $\mu M$  \cite{Gittes1993,McCullough2011}.


\begin{figure}[h]
	\centering
		\includegraphics[width=0.5
        \linewidth]{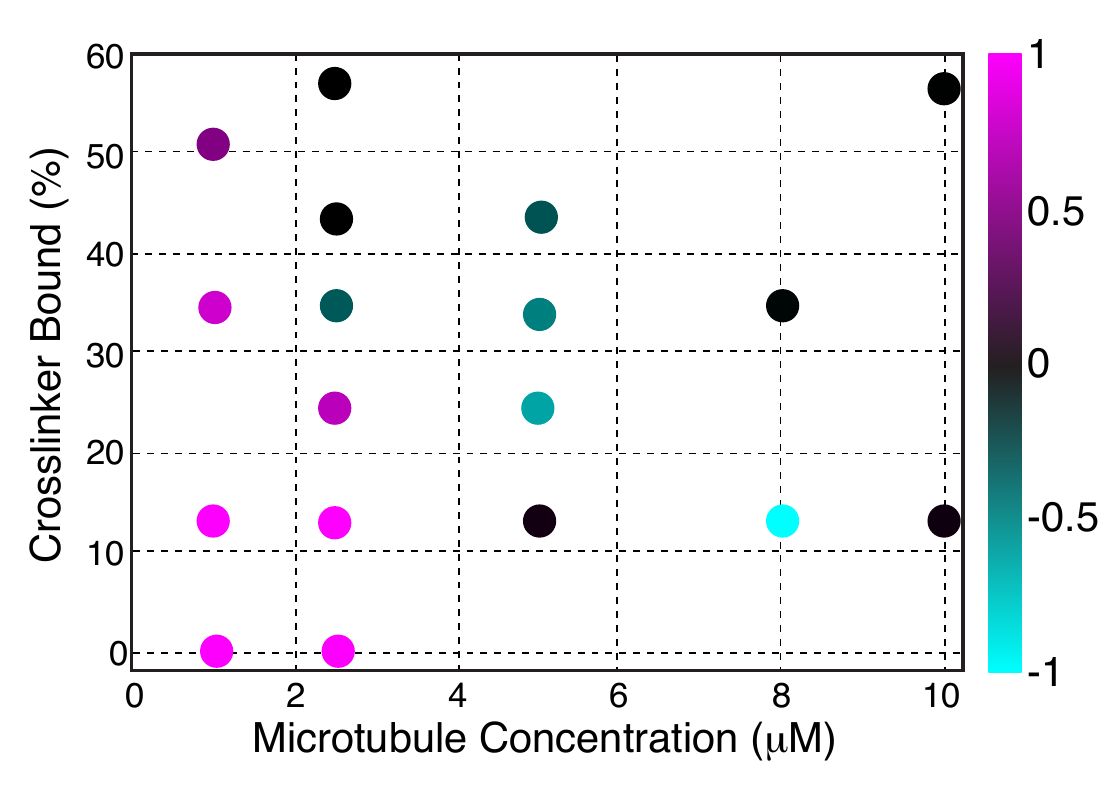}
		\caption{State diagram of active network dynamics. X axis here is microtubule concentration whereas Y axis is percentage of crosslinker bound. The dynamics is color coded from the maximum rate of spreading (magenta, normalized to 1) to the maximum rate of contraction (cyan, normalized to -1). The intensity of the color signifies the relative value of spreading or contraction. We can see the existence of static phases not just at high microtubule density or high crosslinker concentration but also at intermediate densities (microtubule concentration is 5 $\mu M$ MAP65 concentration is 13\%). We also see that all contractile phases are concentrated in a small band like region.
}
	\label{Phasefig}
\end{figure}

We performed these experiments over a range of microtubule concentrations and crosslinker binding percentages, and repeated the measurement for at least three distinct networks at each location in parameter space. For each network, we averaged the normalized variance and used that to quantify the spreading rate metric. Combining all the data together into a single state diagram reveals that the types of dynamics (extensile, contractile, or static)  are clustered depending on the concentration of microtubules and crosslinker binding percentage (Fig. \ref{Phasefig}).

In the absence of any crosslinker, all networks of microtubule bundles are extensile for any density of microtubules. We performed these experiments using PEG to bundle microtubules.  At low concentrations of microtubules and percentages of crosslinkers, the networks of bundles were extensile. The dynamics of such extensile networks was being driven by the activity of microtubules with the kinesin-1 motors on the surface. The crosslinkers most likely work to set-up local bundles of antiparallel microtubules. Thus, at the edges of the networks of bundles, there should be just as many filaments pointing outward as inward. If there was not enough cohesion between microtubules, these filaments would be driven according to their polarity - causing an overall expansion of the edges. We observed that the spreading rate decreases with increasing filaments or crosslinkers (Fig. \ref{Phasefig}, magenta), implying that the crosslinking acts to slow the kinesin-1 driven dynamics. As we know that crosslinkers only act locally between neighboring microtubules, the crosslinkers act as a local cohesion between these bundles of microtubules.

At the other extreme, very high concentrations of microtubules or crosslinkers, the networks are static. Since the crosslinker density is fixed, it is possible that slowing down of dynamics is due to local filament density only. Prior work has shown that dynamics of rod-like particles slow down as the volume fraction increases \cite{Doi1999,Yadav2012}. In our experiments, the microtubule concentration was varied from $0.5~ \mu M$ to $10~ \mu M$. These values correspond to volume fractions, $\phi$, ranging from approximately $10^{-4}$ to $10^{-3}$. Even for these small values of $\phi$, the dynamics can slow down significantly due to the large aspect ratio of microtubules \cite{Doi1999}. Since an increase in volume fraction leads to a loss of rotational and translational degrees of freedom, we would expect a purely monotonic decrease of spreading speed with microtubule density. 

\begin{figure}
	\centering
	  \includegraphics[width=0.4\linewidth]{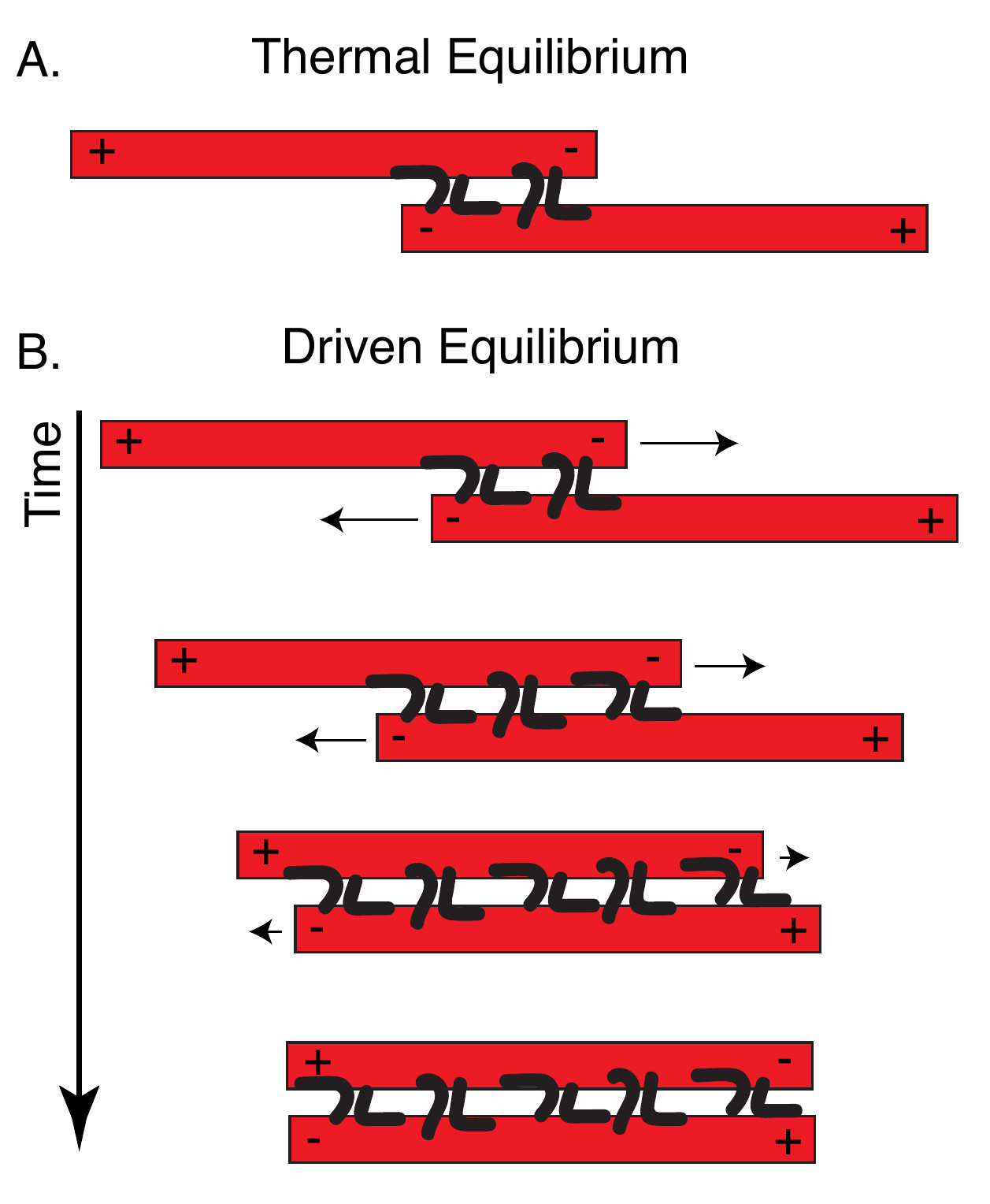}
		\caption{Cartoon demonstrating the suspected mechanism of contraction. (A) Bundles are formed through thermal equilibration of microtubules with crosslinkers, but the overlap is unlikely to be optimal. (B) When the bundle lands on the kinesin, it will be driven to overlap.}
	\label{CartoonFig}
\end{figure}

We observe a non-monotonic dependence implying that the crosslinker is important for driving the contraction, and not just filament density. This fits well with our previous results that showed that higher concentrations of MAP65 slowed the gliding of microtubules as a function of crosslinker percentage \cite{Pringle2013}. We also previously showed that the most likely mechanism for the complete halting of microtubules is from antiparallel microtubules passing each other and crosslinking dynamically during gliding. If the amount of crosslinker is high enough, the microtubules stall due to the fact that they are being pulled in opposite directions. In another set of studies, we examined the effect of MAP65 on single molecule kinesin-1 motility \cite{Conway2014-1,Conway2014-2}. In these studies, we showed that MAP65 does not affect the association time of individual kinesin-1s, but can slow down single kinesin-1 motors \cite{Conway2014-1}. Further, bundling with MAP65 actually enhances the association time of single kinesin-1 motors because the bundle offers more sites for kinesin-1 motor to bind \cite{Conway2014-2}. Taken together, our current and prior work indicates that the static bundles are static because crosslinkers are holding  microtubules tightly together. The motors are trying to push the filaments in opposite directions, but are stalled due to high cohesion between oppositely oriented filaments (Fig. \ref{CartoonFig}). Thus, the ``activity'' of static networks is likely completely controlled by the crosslinking of microtubules into bundles locally.

At intermediate concentrations between these two extremes, however, we observe a prominent, albeit narrow, contractile regime. In the contractile regime, we never observe buckling of microtubules, but we do see rearrangements of individual filaments within the network. We postulate that the rearrangements result in an increase in microtubule overlap at the local, bundle level. The mechanism for these rearrangements is that individual microtubules, which are thermally equilibrated before being added to the kinesin-1 gliding assays, are not maximally overlapped (Fig. \ref{CartoonFig}A). When put in contact with the driving kinesin, these filaments will first move together to increase the region of overlap. The increasing overlap will increase the cohesion of the MAP65, as we previously observed between two microtubules \cite{Pringle2013}. The increased cohesion will eventually stall the motors trying to move the microtubules in opposite directions and cause the filaments to reach a new, motor-driven equilibrium position, which is a shorter bundle, locally (Fig. \ref{CartoonFig}B). Performing this same driven equilibration over the entire network will cause the entire network to contract. 

Our contractile phase is distinct from those reported previously for actin-myosin or other microtubule systems. For actin-myosin systems, prior work showed that actin networks can contract, but the contraction mechanism depends on actin filament buckling \cite{Murrell2012}. In contrast, our mechanism, from both experiments and simulations, is overlap rearrangement. Such rearrangement is likely impossible if the crosslinker is strong or static. Our self-organization is likely made possible because the MAP65 proteins are weak, transient crosslinkers \cite{Pringle2013}. Prior work on microtubule-kinesin systems that contract showed contraction stemming from the creation and mobility of microtubule asters, whereas microtubule bundles were always extensile \cite{Needleman2015,Sanchez2012}. Our mechanism is distinct from those previously studied. Further, unlike all prior studies, our system has the ability to be tuned from extensile to contractile by changing the crosslinker or filament density. 

\subsection*{Simulations}
In order to test the proposed mechanism behind the active network dynamics we observe (Fig. \ref{CartoonFig}), we performed a simple simulation on one-dimensional bundles as model networks. As we describe above, all the activity of the networks can be simplified to the activity of the individual bundles because the crosslinkers only work on neighboring microtubules. Further, MAP65  can crosslink microtubules at different angles, but have a preference for shallow angles and these angles evolve toward antiparallel alignment \cite{Tulin2012}.

Based on these previous observations, we performed a simplified simulation of a one-dimensional ``network'' of microtubules. We initiated the network in a random configuration. The location of the center of mass of each filament along one dimension and the orientation of the filament polarity, which dictates the direction of gliding, were initially randomized. This initial bundle represented the network of bundles that reached equilibrium thermally. 

In the absence of crosslinkers, we assumed the speed of self-propelled filaments, $V_{free}$, was constant. Our previous observations showed that the binding of crosslinkers is cooperative and that crosslinking slowed down antiparallel microtubules that overlapped each other \cite{Pringle2013}.  Thus, the simulation dynamics depended on only the same two variables that modified the velocity: the microtubule density and the fraction of crosslinkers bound to each microtubule. Finally, we assumed that all microtubules have a fixed length $l$ to simplify the model.


\begin{figure}
	\centering
	  \includegraphics[width=0.5\linewidth]{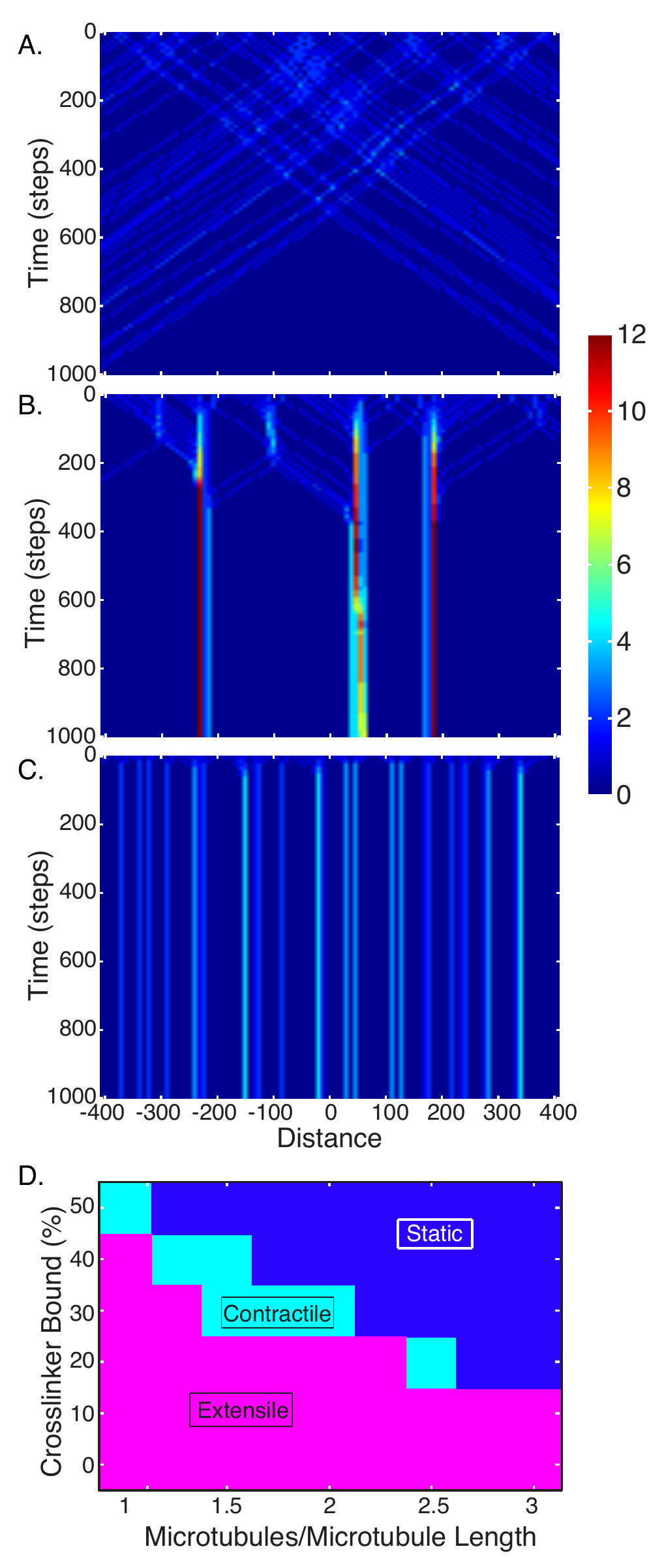}
		\caption{Results from simulations. (A) Example kymograph of extensile activity. (B) Example kymograph of a contractile activity. (C) Example kymograph of a static activity. For each kymograph, the x-axis represents the location as a function of time. The y-axis is the time in simulation steps. The intensity scale represents the number of microtubules. (D) The activity state diagram obtained from our model. The final state of the network was determined by the number of microtubules left inside the system. Contractile and static states were differentiated by comparing the mean distance traveled by a microtubule before it came to rest (for details refer SI). The simple one-dimensional model qualitatively captures the dynamics shown in experiments as shown in Fig.\ref{fig:Trackfig}.}
	\label{fig:SimulatedPhasefig}
\end{figure}

We formulated an expression to determine the speed of the microtubule in the presence of overlapping, antiparallel microtubules as a function of the local density of microtubules, $\rho_{MT}$, the average density of crosslinkers, and the polarity. The crosslinker density was parameterized as the fraction $f$ of crosslinkers bound to single microtubules, which is directly proportional to the percentage of MAP65 bound, as we use in our experiments. The fraction $f$ exists from 0 to 1, but in simulations $f$ was varied from 0 to 0.55 to mimic our experiments. The speed decrease arose from both microtubule crowding as well crosslinker binding between pairs of nearby, antiparallel microtubules. The crosslinker cooperativity that we observed in our previous publication \cite{Pringle2013} suggests that the decay of activity should be more sensitive to the change in crosslinker concentration than to the density of microtubules.

We modeled these effects by assuming the microtubule velocity was a product of two linear decay terms, one of which depended on microtubule density and the other on crosslinker concentration,
\begin{equation}
V_i = V_{free}(1 - \frac{\rho_{MT}}{\rho_{max}})(1-f\frac{\sum_{-l}^{+l} O_i}{O_{max}}),
\label{Simulation}
\end{equation}
where $V_i$ was speed of $i^{th}$ microtubule, and $\rho_{max}$ and $O_{max}$ are the maximum density and total overlap respectively at and above which all dynamics halt. To define the overlap, $O$, we noted that, for an arbitrary microtubule with its center of mass at $x$ pointing along positive x axis, the only other microtubules it can crosslink with are located within $(x-l,x+l)$. For a neighboring microtubule located at position $x'$, we defined its overlap $O$ with the microtubule at $x$ as
\begin{equation}
 O =
  \begin{cases} 
      \hfill l-\left|x-x'\right|    \hfill & \text{ if $x-l<x'<x+l$} \\
      \hfill 0 \hfill & \text{ otherwise} \\
  \end{cases}.
\end{equation}
Thus, $\sum_{-l}^{+l}O_i$ in Eq. (\ref{Simulation}) is the total overlap of $i^{th}$ microtubule at $x$ with its anti-parallel, aligned neighbors in the range $(x-l,x+l)$. F Complete details of simulation parameters can be found in the supplemental information (Supp. Fig. S2).

We simulated a system of $N$ microtubules of length $l$ randomly oriented in one dimension, with the transport direction either toward the right or left along the x-axis, evolving according to equation (\ref{Simulation}). The simulations ran for an order of magnitude longer than the time it takes a free microtubule to travel through the system size. Each sample corresponded to a different value of $\rho_{MT}$ and $f$ and was repeated $100$ times to generate statistics on the type of activity observed. We tracked the number of microtubules within the system and the mean distance traveled by each microtubule.

As we varied $\rho_{MT}$ and $f$ in our simulation, we identified dynamic states of networks as either extensile, contractile, or static by comparing the mean distance traveled by the microtubules and the final number of microtubules inside the system (Fig. \ref{fig:SimulatedPhasefig}A-C).  In our simulation, both the contractile and static states were ultimately static at long times. In order to differentiate between the contractile and the static states, we measured the mean distance traveled by the filaments. Microtubules in the static state move little (less than half the body length) 
before coming to rest and do not show a significant density change (Fig. \ref{fig:SimulatedPhasefig}C). On the other hand, microtubules in contractile states travel significantly longer distances (more than half the body length of filament) before becoming trapped in a few high density locations (Fig. \ref{fig:SimulatedPhasefig}B). Using these definitions of static and contraction were reasonable given our experimental data (Fig. \ref{fig:Trackfig}). As clearly observed in the kymographs, the static networks did not change size or density with time (Fig. \ref{fig:Trackfig}Ciii). Small rearrangements at the beginning of the dynamics do occur before the network froze into place (Fig. \ref{fig:Trackfig}Cii). Contracting networks, on the other hand, not only became smaller, but also had an increase in the intensity, signaling increased filament density in that region (Fig. \ref{fig:Trackfig}B). 

The resulting phase space plot has excellent qualitative agreement with our experimentally obtained phase space (Figs. \ref{fig:SimulatedPhasefig}D, \ref{Phasefig}). Prior theoretical and simulation work studying the effect of alignment-induced interactions among discrete particles has been discussed through the Vicsek model, or the ``flying XY'' model \cite{Vicsek1995,Kosterlitz1973,Tu1995}. Similar problems have also been studied using a continuum approach \cite{Kruse2004,Kruse2005,Bertin2009,Farrell2012}. A broad theme in these works is how local interaction leads to global alignment. In our simulations, we take the route of a simple, 1D model of interacting filaments. We do not require more complicated, 2D models because we started with the experimentally determined assumption that our crosslinkers align and bundle filaments \cite{Pringle2013}. Future models of similar systems that allow the initial evolution of the microtubules into networks could be of interest to theorists. Systems consisting of filaments, interacting in 2D through polymerization or motor proteins will be interesting as well and can take advantage of prior experimental results \cite{Tulin2012,Pringle2013}.

As the model makes clear, the contraction of bundles can arise entirely from an overlap-dependent slowing down of nearby microtubules, mediated by the crosslinkers, and not from specific contractile forces induced by the crosslinkers or motors themselves.
At very high densities of crosslinker, only a little overlap between neighboring microtubules is required to overcome the forces applied by motors and hence the system comes to a complete halt without showing any significant contraction.

Similarly, as the concentration of microtubules increases at fixed crosslinker density, so does the fraction of antiparallel aligned microtubules. This promotes a higher probability for a crosslinker to bind, thereby joining two microtubules. Above a certain concentration there is sufficient overlap of antiparallel, aligned microtubules and also a sufficient density of crosslinkers to bind at these sites and cause contraction. If we keep increasing the microtubule density, however, the dynamics of the system slows down due to crowding and goes to zero above a specific density. This causes slowing down of contraction and finally ceases any activity.

\subsection*{Cell-Like Networks}
\begin{figure}
	\centering
		\includegraphics[width=1\linewidth]{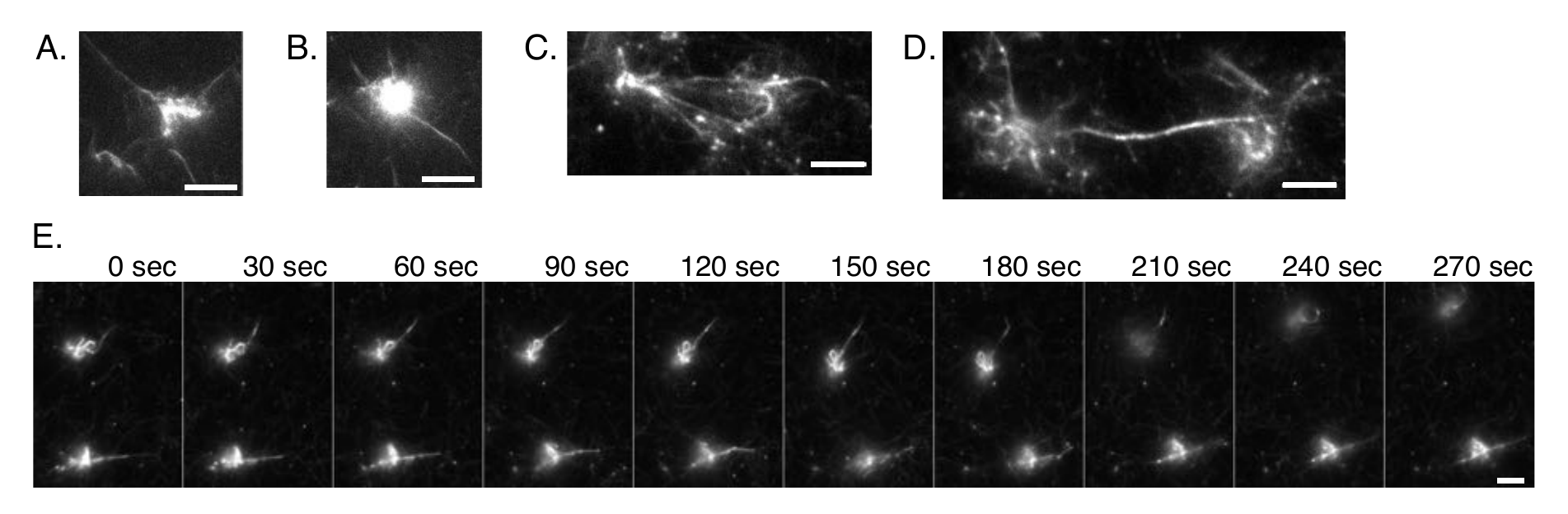}
		\caption{Examples of cell-like networks of microtubule bundles. (A) Networks of bundles with $2.5\mu M$ microtubules and 31 \% MAP65. (B) Networks of bundles with $2.5\mu M$ microtubules and 31 \% MAP65. (C) Networks of bundles with $4.5\mu M$ microtubules and 22 \% MAP65 from previous work \cite{Pringle2013}. (D) Networks of bundles with $4.5\mu M$ microtubules and 22 \% MAP65 from previous work \cite{Pringle2013}.  (E) Time series of dynamics of the networks showing that they can crawl around the cover slip. The top-most network dislodges from the surface during gliding. Networks of bundles with $4.5\mu M$ microtubules and 22 \% MAP65 from previous work \cite{Pringle2013}. Scale bars are 10 $\mu m$ for all images.}
	\label{CellLikeFig}
\end{figure}

In our simulations, the contractile state always ended in a static state where the density of the stationary, high density regions was fixed in time. In the experiments, we sometimes saw the contractile state result in a static state over the long time, although many are still contracting at 45 minutes (Supp. Fig. 2). On other occasions, initially contracting networks would eventually make it past each other and spread apart to become extensile at very long times. Perhaps more interesting, some initially contracting networks would dislodge from the surface to move around as an intact network (Fig. \ref{CellLikeFig}). We previously observed the cell-like networks at one particular concentration of microtubules and MAP65 \cite{Pringle2013}. In most cases, there was a high density ``nucleus'' of filaments with long bundles protruding from them. In many cases the bundle would ultimately be pulled away from the nucleus. 

From our current work, we deduce that these cell-like networks form when the MAP65 is at a sweet spot for cohesion. Further, the fact that we only observe these cell-like networks for small networks of bundles implies that there may be some cut-off size beyond which the networks only extend or become static, but cannot become mobile. These networks were typically initially smaller. At very long times, the dynamics ceased because the network would lose contact with the kinesin-coated surface (Fig. \ref{CellLikeFig}E. The network would remain self-assembled, but without the contact of the motors, it was unable to move in a directed manner.    

\section*{Conclusion}
In this paper, we showed that the dynamics of active crosslinked networks of microtubules are tuned by both microtubule and crosslinker density. We have shown previously that MAP65 slows down antiparallel aligned microtubules in gliding assays in an overlap dependent manner \cite{Pringle2013}. Our experimental results combined with our simulations show that crosslinking of microtubules is essential for contraction. The mechanism behind contraction is the slowing down of microtubule velocity by crosslinkers. The crosslinker induced drag force depends on the bound crosslinker density. This leads to local contraction of bundles and alignment of neighboring microtubules leading to further contraction.

It is essential to understand how various components of a cell interact and produce the dynamics that are hallmarks of all living, cellular systems. To that end, we need to observe and quantify interactions between filaments, motor proteins, and crosslinkers and model those interactions in understanding cytoskeletal mechanics. Towards this aim, we designed an experiment to study the behavior of active microtubule networks in which individual microtubules interact through the action of molecular motors and crosslinkers. We systematically varied the concentration of both microtubules and crosslinkers independently to explore the effect on the dynamics of a network. We showed that, for the first time, an active system of cytoskeletal filaments being driven by motor activity can be tuned through the control parameters of relative density of filaments and crosslinkers. Our system goes through a novel, contractile phase in which networks shrink instead of expanding.

To elucidate the mechanism of these dynamics, we simulated a microtubule bundle as a collection of active rods in one dimension interacting via density driven dynamics. Using our simple model, we were able to recapitulate extensile, contractile, and static dynamic phases and generate a dynamic phase diagram that is qualitatively similar to experimentally obtained data. Our work suggests that the cell can tune its dynamics through tuning the relative density of filaments and crosslinkers to create phases that are biologically important such as the mitotic spindle and contractile ring. Understanding the fundamental principles behind such systems will elucidate how cells autonomously sense, compute, and respond to their environment through shape changes of the cytoskeleton.

\section*{Acknowledgements}
We would like to thank Dr. Peker Milas and Dr. Art Evans for constructive conversation about particle tracking, active matter theory, and models. K.S. was funded by NSF Inspire grant 1344203 to J.L.R, NIH RO1-GM109909 to J.L.R. and The Mathers Foundation. V.Y. was funded by NSF Inspire to J.L.R., MURI 67455-CH-MUR, and Mathers Foundation. C.S. is funded by NSF EFRI ODISSEI 1240441 and Keck Foundation. J.L.R. was funded by NSF INSPIRE, Mathers Foundation, NIH, and MURI.

\bibliographystyle{unsrt}

\end{document}